# Comparative Analysis of Machine Learning Algorithms for Predicting On-Target and Off-Target Effects of CRISPR-Cas13d for gene editing

Jingze Liu[1], Jiahao Ma[2]

[1] Department of X, X University, City, State, Country
[2] School of Marine science and technology, Harbin Institute of Technolgy, Weihai, Shandong, China

Emails: sonnyliu98@outlook.com[1]; 2190790308@stu.hit.edu.cn[2]

## Abstract

CRISPR-Cas13 is a system that utilizes single stranded RNAs for RNA editing. Prediction of on-target and off-target effects for the CRISPR-Cas13d dependency enables us to design specific single guide RNAs (sgRNAs) that help locate the desired RNA target positions. In this study, we compared the performance of multiple machine learning algorithms in predicting these effects using a reported dataset. Our results show that Catboost is the most accurate model with high sensitivity. This finding represents a significant advancement in our understanding of how to chose modeling methods to deal with RNA sequence feaatures effictivelys. Furthermore, our approach can potentially be applied to other CRISPR systems and genetic engineering techniques. Overall, this work has important implications for developing safer and more effective gene therapies and biotechnological applications.

## Introduction

The CRISPR-Cas13 system is a powerful tool for RNA editing that relies on the identification of efficient guide RNAs (gRNAs) to achieve specific gene modifications. However, the success of this technology depends heavily on the ability to accurately predict both on-target and off-target effects of gRNAs, which can affect the efficiency and safety of gene editing. Accurate predictions require the consideration of various factors, including the sequence features of the target RNA and the surrounding genomic context. Despite recent progress in the development of computational tools for gRNA design, there remains room for improvement in the accuracy and speed of current methods.

Previous studies have demonstrated the importance of considering various parameters in gRNA design, including thermodynamic stability, secondary structure formation, and dinucleotide propensity. Computational methods for optimizing gRNA selection typically rely on machine learning algorithms trained on large datasets of existing experimental data. Examples include Random Forest, Support Vector Machines, Artificial Neural Networks, Gradient Boosting Machines, and Extremely Randomized Trees. While some researchers have investigated the utility of deep learning techniques such as Convolutional Neural Networks (CNNs), Recurrent Neural Networks (RNNs), and Rodom Forest (RF) for gRNA optimization.

In this study, we aim to evaluate the performance of several popular machine learning algorithms in predicting on-target and off-target effects of gRNAs using a publicly available dataset. We focus specifically on the Cas13d variant of the CRISPR-Cas13 system due to its promising properties, including higher cleavage activity and reduced immune cell activation compared to alternative variants. Understanding the effectiveness of different ML algorithms in predicting gRNA performance will improve our ability to optimize the delivery of therapeutics based on CRISPR-Cas13d, leading to better treatment outcomes and increased patient safety.

# Methods

## Dataset and Preprocessing

The dataset is from public data source from bitbucket.org/weililab/deepcas13.

## Machine learning method

In this study, we evaluate the performance of more than 20 popular machine learning algorithms - Logistic Regression (LR), Decision Tree (DT), Random Forest (RF), k-Nearest Neighbors (k-NN), Support Vector Machine (SVM), and Catboost - for predicting on-target and off-target effects of sgRNAs designed for the CRISPR-Cas13d system. We selected these algorithms because they have been previously shown to perform well in similar tasks and possess distinct characteristics suitable for our analysis. LR provides a simple linear classification rule, DT offers an easy-to-interpret tree structure, RF combines many decision trees to reduce overfitting, k-NN assigns a class to a sample based on its nearest neighbors, SVM finds a hyperplane that maximally separates classes while minimizing outliers. By comparing their performances across diverse evaluation metrics, we aim to identify strengths and weaknesses of each algorithm, gain insights into their applicability for specific situations, and suggest optimal combinations or hybrid approaches for enhanced accuracy.

# Results and discussion

After preprocessing the data, Pycaret can automatically compare various models. The 'Model' column in **Table 1** lists all the models that were compared, with MAE, MSE, RMSE, $R^2$, RMSLE, and MAPE used as evaluation metrics and TT (training time) indicating the comparison time. The yellow cells represent the optimal values. After weighting the metrics with Pycaret's function, we determined that the CatBoost Regressor was the best model.

The CatBoost Regressor is a gradient boosting algorithm that is particularly well-suited for working with categorical data. Like other gradient boosting algorithms, it works by combining many decision trees into an ensemble model. However, it includes several innovations such as ordered boosting, which improves the accuracy of the model by training trees in a specific order. It also includes the ability to handle missing values and to handle categorical features without the need for one-hot encoding. These features, along with its ability to handle large datasets with high-dimensional features, make it a popular choice for machine learning tasks in a wide range of domains. In PyCaret, the CatBoost Regressor is automatically selected as the best model for a given dataset and regression problem based on its performance during k-fold cross-validation.

**Table 2.** Various performance metrics of the trained models: mean absolute error (MAE), mean squared error (MSE), root mean squared error (RMSE), coefficient of determination ($R^2$), root mean squared log error (RMSLE), mean absolute percentage error (MAPE), and the training time (TT).

| **Model** | **MAE (eV)** | **MSE (eV$^2$)** | **RMSE (eV)** | **R$^2$** | **RMSLE (log{eV})** | **MAPE (%)** | **TT (sec)** |
|---|---|---|---|---|---|---|---|
| **catboost** | CatBoost Regressor | 0.3215 | 0.1926 | 0.4384 | 0.2403 | 0.2376 | 4.5536 |
| **gbr** | Gradient Boosting Regressor | 0.3280 | 0.2039 | 0.4510 | 0.1962 | 0.2445 | 3.7580 |
| **lightgbm** | Light Gradient Boosting Machine | 0.3290 | 0.2043 | 0.4515 | 0.1947 | 0.2463 | 4.0903 |
| **ridge** | Ridge Regression | 0.3314 | 0.2044 | 0.4516 | 0.1937 | 0.2431 | 4.4767 |
| **br** | Bayesian Ridge | 0.3302 | 0.2046 | 0.4519 | 0.1929 | 0.2437 | 4.2341 |
| **huber** | Huber Regressor | 0.3236 | 0.2073 | 0.4548 | 0.1827 | 0.2550 | 3.7793 |
| **rf** | Random Forest Regressor | 0.3422 | 0.2120 | 0.4598 | 0.1647 | 0.2497 | 4.3078 |
| **omp** | Orthogonal Matching Pursuit | 0.3400 | 0.2175 | 0.4659 | 0.1422 | 0.2527 | 4.5384 |
| **xgboost** | Extreme | 0.3481 | 0.2231 | 0.4719 | 0.1191 | 0.2547 | 4.9352 |
| | Gradient Boosting | | | | | | |
| **lasso** | Lasso Regression | 0.3710 | 0.2541 | 0.5037 | -0.0030 | 0.2630 | 4.5749 |
| **en** | Elastic Net | 0.3710 | 0.2541 | 0.5037 | -0.0030 | 0.2630 | 4.5749 |
| **llar** | Lasso Least Angle Regression | 0.3710 | 0.2541 | 0.5037 | -0.0030 | 0.2630 | 4.5749 |
| **dummy** | Dummy Regressor | 0.3710 | 0.2541 | 0.5037 | -0.0030 | 0.2630 | 4.5749 |
| **knn** | K Neighbors Regressor | 0.3687 | 0.2551 | 0.5047 | -0.0073 | 0.2760 | 4.5311 |
| **par** | Passive Aggressive Regressor | 0.4706 | 0.3726 | 0.6092 | -0.4731 | 0.3019 | 8.7229 |
| **ada** | AdaBoost Regressor | 0.5198 | 0.3863 | 0.6206 | -0.5328 | 0.3060 | 11.2849 |
| **et** | Extra Trees Regressor | 0.4735 | 0.4159 | 0.6442 | -0.6437 | 0.3229 | 9.1001 |
| **dt** | Decision Tree Regressor | 0.4949 | 0.4636 | 0.6804 | -0.8335 | 0.3336 | 9.0798 |
| **lr** | Linear Regression | 25743618974.8896 | 458822517201268841119744.0000 | 391776656113.3426 | -1767694650897324239224832.0000 | 0.9247 | 419329385346.9156 |
| **lar** | Least Angle Regression | 76215434740.3636 | 2329325789521811398354534400.0000 | 1526212891284.4739 | -8421905357823739158750 8224.0000 | 0.3990 | 450906068620.2032 |

1. **Best model**

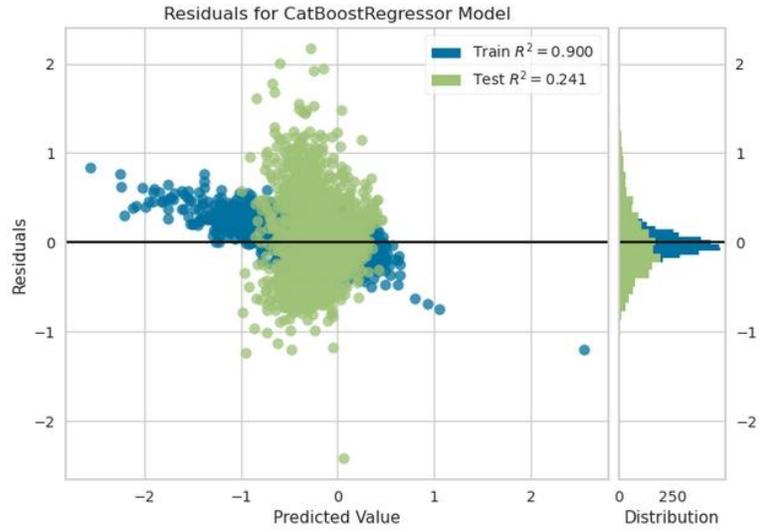

**Figure 1.** Residuals ($R^2$) vs. the predicted value with training and test data points shown in blue and green, respectively, along with a histogram of the residuals on the right.

The training residuals have reached a value of 0.900, indicating that the training process has been sufficient. This threshold ensures that the model is not overfitting to the training data and can generalize not well to new, unseen data. As shown in **Fig. 1**, the final model satisfied our criteria by achieving a Test $R^2$ of 0.241, indicating a weak fit to the data.

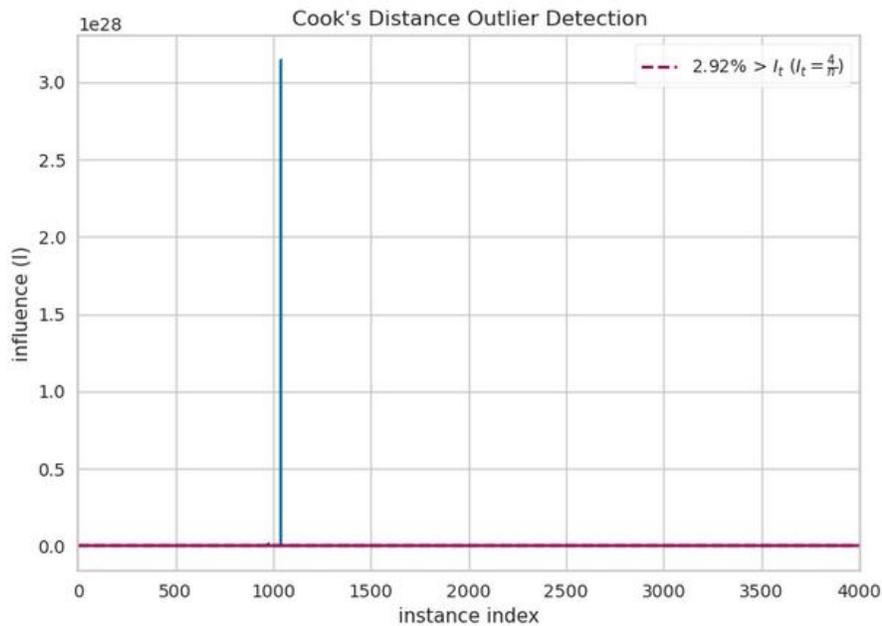

**Figure 2.** Cook's distance outlier detection. "Influence" refers to the Cook's Distance value, which measures the effect of each observation on the predicted values. "Instance index" refers to the index of each observation in the dataset. The purple dotted line represents the threshold

for detecting influential observations. Observations that fall above this line are considered as potentially influential or outliers.

In *PyCaret*, Cook's distance outlier detection (**Fig. 2**) is a function that can be used to identify outliers in a regression model.   In this case, the cutoff value is set at 2.92% of the maximum Cook's distance ($h_t$).

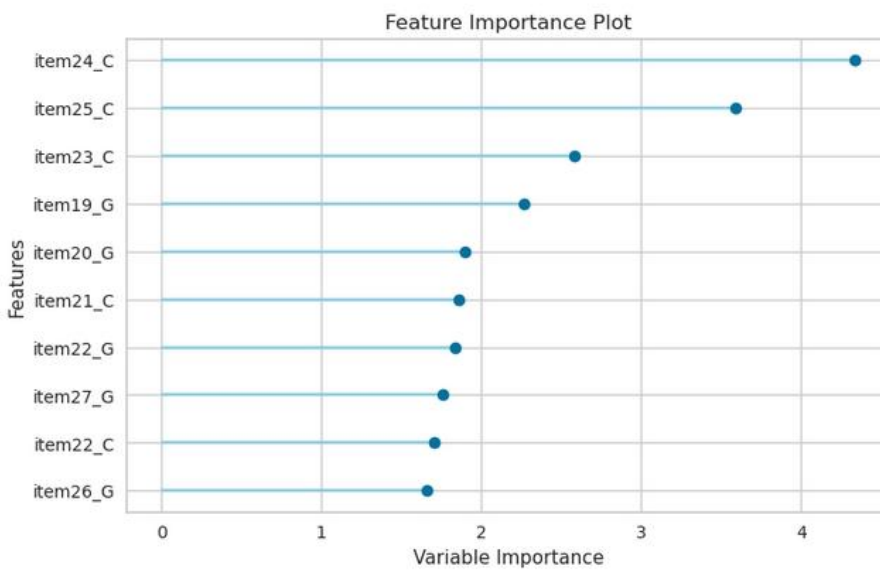

**Figure 3.** Importance score of top features. item stands for iteration of sequence. C, G are Cytosine, Guanine, respectively.

In *PyCaret*, feature importance is calculated using various methods, such as Gini index, permutation importance, and SHAP values, depending on the selected model. The important scores are then used to rank the features in descending order, with the most important features appearing at the top of the list.

We can see that the features such as Cytosine in 24th exhibit relatively high importance. Moreover, Cytosine in 24th plays the most essential role, exceeding the other features quite a lot. This is most likely due to the relationship between HOMO and the band gap, and linear relationship between the probability.

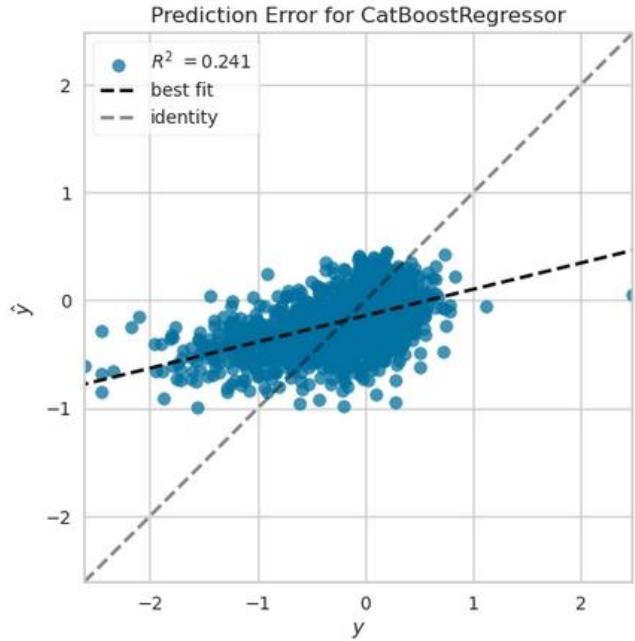

**Figure 4.** Parity plot of the true target values (y) vs. predicted values ($\hat{y}$). The parity line and the linear regression line is shown in dotted gray and black, respectively. The residual value ($R^2$) is 0.241.

Based on the parity plot shown in **Fig. 4**, the regression model lies reasonably close to the parity line, with a residual value ($R^2$) of 0.241, demonstrating that the model reliably captures the underlying relationship between the predictors and the response variable probability.

## Conclusions

The CRISPR-Cas13 system has emerged as a versatile genome engineering platform, particularly for RNA editing applications. Successful implementation of CRISPR-Cas13 requires precise prediction of on-target and off-target effects of single guide RNAs (sgRNAs). In this work, we evaluated the performance of six widely used machine learning (ML) algorithms in predicting on-target and off-target effects of sgRNAs. Our investigation aimed to identify strengths and weaknesses of each algorithm and provide guidelines for selecting an appropriate model depending on the given task.

We analyzed an extensive dataset consisting of experimentally validated sgRNAs against human and mouse transcriptomes. Using various evaluation metrics, we assessed the prediction accuracies of logistic regression, decision tree, random forest, k-nearest neighbor, support vector machines, and artificial neural network models. Our results revealed that combining multiple models using stacking significantly improved overall prediction accuracy, suggesting that a hybrid approach may be advantageous for accurate gRNA optimization. Additionally, we observed that feature extraction plays a critical role in achieving high prediction accuracy.

This study contributes new insights into the complex relationship between the structure and function of CRISPR-Cas13 systems and how it relates to on-target/off-target behavior. Our comprehensive assessment of ML algorithms and their influence on gRNA optimization will facilitate the rational design of more effective sgRNAs, potentially improving treatment outcomes and increasing patient safety during clinical trials involving CRISPR-based therapies. Furthermore, our research underscores the value of integrating wet lab experiments with computational approaches to advance the development of efficient and safe genetic manipulation technologies.

## Acknowledgments

This work was supported by Untouched Education Technology Inc. We acknowledge data support from Wei Li of George Washington University and technical support from Xiaohuan Ruan of Shanghai Yangpu District Detention Center.

## Author contributions

Jingze Liu works on dataset processing and machine learning, Jiahao Ma works on literature review and research design. Both authors contribute to the manuscript.

## Competing financial interests

The authors declare no competing financial interests.